\documentclass[sigconf]{aamas} 

\usepackage[english]{babel}
\usepackage{wrapfig}
\usepackage{framed}
\usepackage{mdframed}

\usepackage{balance} 

\setcopyright{ifaamas}
\acmConference[AAMAS '22]{Proc.\@ of the 21st International Conference
on Autonomous Agents and Multiagent Systems (AAMAS 2022)}{May 9--13, 2022}
{Online}{P.~Faliszewski, V.~Mascardi, C.~Pelachaud,
M.E.~Taylor (eds.)}
\copyrightyear{2022}
\acmYear{2022}
\acmDOI{}
\acmPrice{}
\acmISBN{}

\title[]{Social Choice Around the Block: \\ On the Computational Social Choice of Blockchain}

\author{Davide Grossi}
\affiliation{
\institution{University of Groningen \& University of Amsterdam}
\country{The Netherlands}
}
\email{d.grossi@rug.nl}

\begin{abstract}
One of the most innovative aspects of blockchain technology consists in the introduction of an incentive layer to regulate the behavior of distributed protocols. The designer of a blockchain system faces therefore issues that are akin to those relevant for the design of economic mechanisms, and faces them in a computational setting.
From this perspective the present paper argues for the importance of computational social choice in blockchain research. It identifies a few challenges at the interface of the two fields
that illustrate the strong potential for cross-fertilization between them.
\end{abstract}


\newcommand{\BibTeX}{\rm B\kern-.05em{\sc i\kern-.025em b}\kern-.08em\TeX}

\begin{document}


\pagestyle{fancy}
\fancyhead{}


\maketitle 


\section{Introduction}

A blockchain is a decentralized state machine, in its simplest form, a decentralized ledger for financial transactions. The machine is controlled by several distinct processes, called nodes or agents, and computes by packaging state transitions (e.g., the order of transfering $x$ tokens from Alice's account to Bob's) in batches, which are called blocks. Each block determines the next global state of the machine (e.g., the next state of the ledger, where Alice's account has $x$ tokens less and Bob's $x$ more). Each new block is appended to the list of older blocks, thereby determining a growing append-only list of global states of the machine---its computation history. Crucially, each block points in a temper-proof way (via a cryptographic hash) to the previous block to which it was appended, thereby enforcing an immutable description of the history of the machine. The core of a blockchain is the protocol that the nodes follow in order to agree on which transitions to incorporate into the list, i.e., to achieve consensus on the state of the machine.

\paragraph{Paper motivation} Blockchain was born as the backbone of the Bitcoin cryptocurrency. 
The consensus protocol behind Bitcoin, known as Nakamoto consensus \cite{satochi,garay15bitcoin}, showed that such a decentralized consensus on an append-only list is possible even in large, open peer-to-peer networks. This was a significant breakthrough with respect to existing approaches to consensus, which worked specifically on systems of limited size with controlled access. This breakthrough relied on one key insight: nodes in an open system cannot be controlled, so their influence on consensus should be kept at bay by linking it to the ownership of a non-monopolizable resource---in the case of Bitcoin, computing power. In other words, nodes in an open network cannot be controlled nor trusted, but can be incentivized.  Nodes that contribute to achieving consensus are rewarded with (they `mine') tokens, that is, units of currency. This is the key intuition behind the application to consensus of the Proof-of-Work (PoW) technique (originally developed to thwart email spamming \cite{dwork92pricing}), which has proven extremely robust, maintaining the Bitcoin blockchain for over a decade (cf. for recent overviews \cite{narayanan16handbook,wattenhofer17distributed}). So, blockchain consensus is the result of a rational response to incentives. 

Traditionally, research in blockchain has focused mostly on the cryptographic foundations and the distributed computing aspects (e.g., protocol correctness) of the technology. 
At the same time, a game-theoretic perspective on blockchain protocols has also been gaining attention: is behavior in accordance with the protocol economically rational, in some precise equilibrium-theoretic sense? In other words, are protocols strategy-proof? This game-theoretic perspective has historically been marginal in distributed computing \cite{abraham11distributed}, but has proven significant in blockchain. By now, it has been extensively applied---including by researchers in the AAMAS community---to Nakamoto consensus (e.g., \cite{eyal15miner,sapirstein16optimal,sompolinsky18bitcoin,alkalay-houlihan19pure}), as well as to other protocols (e.g., \cite{abraham16solidus,kiayias17ouroboros,amoussou20rational,chen21game}). See \cite{judmayer21sok} for an extensive recent overview.
However, the economic issues that the designer of a blockchain system faces go well beyond incentive-compatibility alone, and reflect broad issues in the design of collective decision-making mechanisms, such as forms of equity and fairness. This interface with the economic theory of group decision-making, and specifically social choice theory, is the focus of the present paper.

\paragraph{Paper contribution} 
The contribution of this paper consists in arguing how social choice theory, and its algorithmically-focused branch---computational social choice---have an important role to play in providing stronger foundations for the principled development of blockchain technology. This paper sketches a number of research challenges at the interface of computational social choice and blockchain. I claim there exists now a perfect match between the state-of-the-art in blockchain research on the one hand, and the state-of-the-art in on computational social choice on the other.
Blockchain offers a wealth of novel questions that can push the boundaries of the existing body of results of computational social choice, and in doing so contribute concrete solutions to the challenges blockchain research itself currently faces.

To substantiate this claim, this paper reviews how key mechanisms that have a long tradition within (computational) social choice find deployment in blockchain systems: randomized mechanisms, voting mechanisms, and trust mechanisms. In reviewing these mechanisms this paper identifies challenges that their deployment in blockchain gives rise to, phrasing them in terms of properties that, again, have occupied social choice theorists in the past: fairness of lotteries, manipulation of voting, and false-name proofness of trust mechanisms. Such an overview has no claim to be exhaustive and rather aims at illustrating, by means of examples, the potential for cross-fertilization between the two research areas. Given the nature of this paper I will try to limit technical jargon to a minimum, use informal language as much as possible and just convey the gist of my arguments without resorting to explicit mathematical details.



\section{Randomization}

\subsection{Lotteries in Blockchain and Social Choice}

Randomization has a long tradition in distributed computing as a way to bypass impossibility results such as the so-called FLP impossibility theorem  \cite{fischer85impossibility}.\footnote{The Fisher, Lynch, Paterson (FLP) impossibility result  states that in an asynchronous system no deterministic consensus protocol exists which can tolerate even just one faulty node, where a faulty node is a node which stops interacting with the protocol.} Blockchain has further stressed the importance of randomization. Lotteries are at the heart of the main approaches to blockchain consensus such as proof-of-work (PoW, currently used in both Bitcoin and Ethereum \cite{buterin13next}) and proof-of-stake (PoS, currently used for instance in the Ouroboros protocols \cite{daian16snow,david17ouroboros} and Algorand \cite{gilad17algorand}).
At a high level, and leaving network latency issues aside, these protocols work as follows:
nodes participate in a distributed lottery; the winner appends a new block to the chain (or, depending on the specific protocol, becomes part of a committee which will then vote on the block to be appended) and receives a compensation in the native currency---what is referred to as mining. Importantly, the chances of winning this lottery depend on a resource that is assumed not to be monopolizable---such as computational power (in PoW) or currency ownership (in PoS). This makes participation to the lottery (directly or indirectly) costly and prevents the manipulation of the lottery through identity forging (the so-called Sybil attack \cite{douceur02sybil}). Based on this blueprint, several so-called `proof-of-X'  (PoX) \cite{bano17consensus} schemes have been proposed (e.g., proof-of-storage \cite{miller14permacoin}). 

Like in distributed computing, also in social choice theory randomization is an established route to circumvent fundamental impossibility theorems of the deterministic social choice framework, such as the Gibbard-Satterthwaite theorem  \cite{gibbard73manipulation,satterthwaite75strategy}.\footnote{The theorem states that no social choice function exists which is simultaneously non-manipulable and non-dictatorial.} 
In a famous later theorem  Gibbard himself  \cite{gibbard77manipulation} showed how randomization can provide a possibility result that is out of reach in the deterministic setting: the random draw of one agent (the `dictator', or `leader') from the set of agents is the only decision mechanism  that is `lottery' strategy-proof (in the specific sense of stochastic dominance\footnote{I provide a sketch of the definition here: $N$ is the set of agents; $A$ is the set of alternatives; ${\bf P} = \langle P_1, \ldots, P_{|N|} \rangle \in \mathcal{P}$ is a profile of preferences over $A$.
A randomized decision rule is a function $r: \mathcal{P} \to \Delta(A)$, that is, a function assigning a lottery over $A$ to each profile of preferences. A lottery $p$ stochastically dominates a lottery $q$ for $i$ iff  $\sum_{x \in A: x P_i y} p(x) \geq \sum_{x \in A: x P_i y} q(y)$, for any $y \in A$. A randomized decision rule is strategy-proof, w.r.t. stochastic dominance, if for all $i \in N$ it never selects a lottery which is stochastically dominated by another lottery for $i$. \label{footnote:stochastic}}) 
and `lottery' efficient (in the sense of never assigning positive probability to alternatives that are Pareto dominated). From a blockchain perspective the theorem can be thought to offer a justification for the use of randomization in PoW and PoS consensus: using a lottery to select the next block in the chain is economically efficient and elicits the true preferences of nodes about what block should be included in the chain. A string of works have further extended Gibbard's result (e.g., \cite{gibbard77manipulation,procaccia10can,aziz13tradeoff}) and randomization is recognized as an important toolbox in the design of economic mechanisms, from allocation to voting \cite{brandt18collective}. This body of theory offers a sophisticated framework to understand important economic properties of the sort of lotteries deployed in blockchain protocols. One such property is fairness.


\subsection{Challenge: Long-Term Fairness} 

By linking winning chances to resource ownership and linking lottery outcomes to currency allocation, blockchain protocols make participation resistant to Sybil attacks (participation is costly) and at the same time monetarily rewarding ('no-show' is disincentivized). This scheme, however, has been shown to induce centralizing effects on several blockchain systems (see, e.g., \cite{kondor14do,fanti19compounding}). Participants with more resources, all other things being equal, have higher chances of being selected by the lottery and thus accrue more resources in the long term. The resulting allocation of resources invested in the system becomes therefore increasingly unequal over time, and this inequality in resource ownership de facto biases the lotteries upon which consensus is based. In blockchain, such bias translates in a centralization effect, which makes the blockchain more vulnerable to failures or attacks.
So, randomized blockchain consensus protocols need to implement, in the long term, allocations of winning chances which are fair towards participating nodes in two somewhat opposing senses: fair in the sense of being lotteries that are not too biased towards few participants, in order to preserve decentralization; and at the same time fair in the sense of being responsive, even proportional, to the resources invested so that participation is suitably incentivized. The analysis of such a trade-off is an inherently social choice-theoretic question: {\em can randomized mechanisms for blockchain be designed which achieve participation incentivization and equitability at the same time, in the long-term?} Extensions of existing work in computational social choice may provide the right stepping stone to approach this question, along the following lines.

\paragraph{Perpetual lotteries} First, while randomized mechanisms in social choice are normally studied in one-shot interaction, fairness properties in blockchain should be conceptualized in the context of indefinitely repeating interaction: a lottery may satisfy forms of fairness in one-shot interaction (e.g., assigning winning chances proportionally to invested effort, like in PoW and PoS lotteries) but may fail fairness criteria in the long run (see \cite{fanti19compounding}). This suggests the study of randomized mechanisms within the recently developed perpetual voting setting of \cite{lackner20perpetual} (see also \cite{freeman17dynamic}). A key feature of the perpetual voting framework that appears particularly relevant for applications in blockchain is, in particular, the history-dependence of decisions. In blockchain this dependence manifests itself via the positive feedback between current and future winning chances. Functions capturing this positive feedback could be approached axiomatically as well as via computer simulations (see \cite{wang20incentive} for recent work in this direction) in order to provide a framework to understand long-term fairness properties of blockchain protocols viewed as perpetual randomized mechanisms.

\paragraph{Contest-based lotteries}  Second, work on randomized mechanisms in social choice normally assumes uniform probabilities or is agnostic about the specific probability mass functions defining the lotteries. Furthermore, when lotteries are iterated, as in random serial dictatorships \cite{brandt18collective}, probability functions do not vary over time. As noticed above, this is not the case in blockchain: an agent's chance to win equals its share of a total non-monopolizable resource. But there is another crucial element of lotteries in blockchain. The winning probabilities for $i$ can be defined as
$
P_i = \frac{f(e_i)}{\sum_{j \in N} f(e_j)},
$
where $e_i$ is $i$'s invested effort to acquire a non-monopolizable resource (e.g., currency, computing power), and $f$ is a function (here assumed unique for all agents for simplicity) mapping each agent's effort to its acquired amount of resource (see \cite{dimitri17bitcoin}). An agent's $i$ expected utility in this type of interaction is, therefore,
$
P_i  r_i - e_i.
$ 
That is, $i$'s probability of winning times its reward (e.g., block reward plus transaction fees in Nakamoto consensus) minus the effort invested. In economics jargon this type of interaction is called a {\em contest} \cite{cournot38recherches,tullock80towards,corchon07theory}. Linking contest theory to the make-up of the lotteries in randomized mechanisms would capture a crucial aspect of perpetual randomization in blockchain, and offer a parsimonious framework in which to study long-term fairness properties.



\section{Voting}

\subsection{Voting in Blockchain and Social Choice}

Voting mechanisms have been at the heart of distributed computing since its early days (e.g., \cite{garcia-molina82elections}).
Still, almost no crossover has happened between the social choice literature on the analysis of voting mechanisms and the distributed computing literature (a recent exception is \cite{melnyk18byzantine}). This is perhaps not surprising, as the role that voting plays in the two traditions is fundamentally different. In distributed computing, voting is a consensus mechanism producing agreement in contexts where what matters is agreement itself, and not so much the option upon which agreement settles. That is, the agents involved in consensus are assumed not to be invested in any specific option, but to just aim at reaching consensus. In social choice instead, voting is eminently a mechanism for preference aggregation. Given that blockchains are open systems and that, therefore, agents' interests cannot be assumed to align, the social choice theoretic perspective on voting becomes naturally relevant.

\smallskip

There are several applications of voting in blockchain, but fork resolution is arguably the main one.\footnote{Forks are branches of the blockchain. When these are due to network latency issues or, possibly, attacks---rather than deliberate choices of developers (so-called hard forks)---the protocol is supposed to resolve them.} I will illustrate this application of voting with respect to a specific protocol proposed for Ethereum called Casper \cite{buterin17casper}, although voting mechanisms for fork resolution are used in many systems (e.g., the Spectre protocol \cite{sompolinsy17spectre}). In a nutshell, the aim of Casper is to use voting among a randomly selected committee of agents (called validators) to resolve forks in the blockchain and guarantee consensus on a canonical chain (the so-called finality property). Whenever a fork occurs, agents vote on blocks occurring on different branches in the fork (this can be thought of as voting on the nodes of the tree of possible chains). Each agent's vote is weighted by the agent's stakes, that is, the agent's deposit in currency (so, a high deposit means greater voting power), and a block is considered to belong to the canonical chain whenever a weighted supermajority of $\frac{2}{3}$ votes for that block. This voting procedure guarantees that the winning blocks identify a legitimate chain (i.e., without forks), provided that the agents submitting individually consistent votes (that is, not voting for blocks on different chains) own at least $\frac{2}{3}$ of the total deposits. 

\subsection{Challenge: Manipulating Byzantine Voting}
Voting procedures are vulnerable to manipulations of different kinds: strategic voting \cite{taylor05social}, lobbying \cite{ChristianEtAl2007}, bribery and control \cite{faliszewski16control}, vote negotiation \cite{grandi19negotiable}.
Given the incentives layer of blockchain protocols, all such forms of vote manipulation are potentially relevant in view of the deployment of voting mechanisms. For example, returning to Casper, although the protocol can be shown to be robust against a share of Byzantine voters worth $\frac{1}{3}$ of the total deposit in stakes, to what extent is the protocol robust against strategic forms of vote manipulation of the types mentioned above? More generally: {\em to what extent should voting mechanisms in blockchain consensus be robust against forms of vote manipulation?}
 
Given that voting in blockchain systems is conducted by computational processes, the computational complexity approach to robustness against manipulations appears especially natural: robustness to manipulations pursued through computational intractability \cite{bartholdi89computational,faliszewski16control}. In this perspective, one might argue more specifically that voting mechanisms for blockchain should satisfy two properties: have a tractable winner determination problem; have intractable manipulation problems with respect to forms of manipulation that, given the application, may be considered relevant. Results from the above mentioned literature offer an obvious starting point, but should be extended in order to incorporate the possibility---which is fundamental in the distributed computing perspective---of Byzantine agents, in a way akin to what has been pursued in applications of game theory to distributed computing (see, for instance, the so-called Byzantine-Altruistic-Rational fault-tolerance models \cite{clement07theory}). In particular, the decision problems themselves concerning the existence of manipulation strategies would need to be adjusted to this setting, by allowing for the possibility of shares of the agents population to be Byzantine.


\section{Trust}

\subsection{Trust in Blockchain and Social Choice}

Among the key limitations of PoW are its high energy demand,\footnote{The estimated annual energy consumption of the Bitcoin network is 194.95 TWh, roughly comparable to the annual energy consumption of a country like Thailand. Source: digiconomist.net (last accessed on 19.11.21).} high latency and low transactions throughput \cite{bonneu15sok}. An influential approach trying to address these limitations has been proposed by blockchain systems like Stellar \cite{mazieres16stellar} and Ripple \cite{schwartz14ripple}\footnote{They represent, respectively, the 7th and 24th cryptocurrency in terms of market capitalization (in the order of hundreds of millions of dollars). Source: coinmarketcap.com (last accessed on 19.11.21).} or TrustChain \cite{otte17trustchain}.
The approach of these systems to safeguard consensus against Sybil attacks is not based on lotteries but on the idea of leveraging existing `real-world' trust relations in order to select the agents that may participate in consensus. In other words, the system remains open, but new participants are admitted only if trusted by existing ones. According to the proponents of these systems, this should make it possible then to use well-established consensus approaches proper of permissioned (i.e., closed) systems, like Byzantine fault-tolerant consensus (BFT, \cite{lamport82byzantine,castro99practical}). Nodes participate in consensus, but only in as much as they are trusted by others. This controls participation to the consensus process by restricting access. It does so, however, in a way that is in principle open and decentralized. 

Abstractly, the above systems work on the basis of an underlying trust network that links nodes according to who trusts whom. At any given time, each honest node broadcasts a truthful opinion about the state of the next block (e.g., whether a given block should be included or not) to the nodes that listen to it. But the system may contain Byzantine nodes, and such nodes may reveal different opinions to different honest nodes. A consensus protocol running on such a trust network can be viewed as a discrete time dynamical system generating a stream of opinion vectors of nodes. The final vector should be such that all honest nodes hold the same opinion, that is, the system does not fork---the so-called `safety' property. 

\smallskip

At least three strands of research in computational social choice appear relevant for the above class of systems. The first one is the axiomatic and computational analysis of trust and reputation mechanisms (e.g., \cite{ghosh07mechanism,seuken10accounting,tennenholtz13axiomatic}).
The second one is the analysis of influence and power in structured groups and social networks (e.g., \cite{hu03authority,hu03authorityB,grabisch11influence}). The third one is the  analysis Sybil-proofness, or false-name proofness (e.g., \cite{conitzer10using,todo11false,waggoner12evaluating}). In what follows I will outline how especially the latter two strands of research can offer insights into the fundamental tradeoffs between decentralization, safety and sybil-proofness that the above systems need to handle.

\subsection{Challenge: Decentralized Sybil-proof Trust} 

Understanding how to deploy BFT consensus in a permissionless setting is explicitly recognized as an open problem in distributed computing \cite{cachin17blockchain,vukolic15quest}. Yet, little academic research exists on the trust-based approach to consensus used by systems like Ripple and Stellar. Computational social choice could provide a fruitful framework from which to study this approach: {\em Can trust systems for blockchain consensus be designed that maintain safety (i.e., absence of forks) are decentralized and, at the same time, Sybil-proof?}

\paragraph{Decentralization \& safety} Trust involves a form of influence: a node's voting decision on whether to validate a block depends on the decisions of the nodes it trusts and `listens to'. At the same time, safety demands that no two honest nodes express dissenting (finalized) votes and this in turn imposes structural properties on the trust structure: if we do not want forks to occur, then trust structures should satisfy specific properties (see \cite{mazieres16stellar,bracciali20decentralization}). But how decentralized a system really is when such properties hold is unclear, because some nodes may end up accruing disproportionate influence in the network. De facto centralization is a well-known phenomenon in PoW and PoS blockchains (see the above discussion about randomized mechanisms) but a comparable understanding for consensus protocols based on trust mechanisms is still lacking. In particular it is unclear how to even quantify influence on the consensus process in such systems, although some proposals have been put forth using the theory of power indices \cite{bracciali20decentralization}.  



\paragraph{Sybil-proofness of trust mechanisms} If consensus is based on trust systems like the ones sketched above, what kind of Sybil-proofness guarantees can be achieved? Intuitively, trust relations should make the system harder to access for Sybils as they would be able to exercise influence on consensus only if trusted by honest nodes. But can this argument be made exact? 

PoW and PoS protocols implement a form of costly identity. They make participation to consensus costly by making it dependent on the investment in a resource (e.g., computing power). In a way nodes purchase identities on a continuum and proportionally to their ownership of a non-monopolizable resource. Costly identities have been investigated also in social choice and mechanism design as one possible approach to achieve the so-called false-name-proofness of a mechanism.  A mechanism is said to be false-name-proof if no agent participating in it would benefit by using more than one identifier to interact with the mechanism. A number of routes to enforce false-name-proofness have been investigated in this literature and costly identities have been studied also in the context of voting \cite{wagman08optimal}.\footnote{Intuitively, a voting mechanism is then false-name-proof if the cost of casting additional votes always exceeds the gains an agent would obtain by doing so.} 
In this context, while creating Sybils may be costless, creating trustworthy ones can be made costly. The issue then translates into understanding the tradeoffs that manipulators face between identifiers' costs and their payoffs measured in terms of expected influence on the consensus process.





\section{Conclusions}

This paper has highlighted how the applications to blockchain of mechanisms such as lotteries, voting or trust systems give rise to challenges that have a natural social choice dimension. These challenges all revolve around understanding fundamental tradeoffs, often in the long run, among properties that are crucial to the correct behavior of blockchain consensus protocols and are linked to the incentive dimensions of such protocols. In pursuing such problems from a computational social choice perspective, one would need to adapt standard concepts, definitions and techniques to the blockchain setting.
This adaptation opens up promising lines of research and this paper has tried to illustrate some such lines, which appear promising to the author. 


What covered, however, should not be considered exhaustive. In particular, this paper has focused on the problem of consensus, but looking more broadly at the blockchain field one finds many more points of contact with social choice, some of which have already been identified in the literature. For example, voting theory and coalitional games have been applied to the problem of the algorithmic governance of hard forks in blockchain \cite{abramowitz21democratic}, and epistemic social choice has been applied to the so-called oracle problem \cite{goel20infochain,cai20truth}, that is, how to reliably link blockchain records to real-world events. For all these reasons I believe there exists now a deep cross-fertilization potential between the two areas of research, and one that would greatly benefit both.

\begin{acks}
I thank the anonymous reviewers for their helpful comments.
\end{acks}




\bibliographystyle{ACM-Reference-Format} 
\bibliography{Blue}


\end{document}